\documentclass{PoS}

\usepackage{epsfig}
\usepackage{amssymb}
\usepackage{amsmath}
\usepackage{graphicx}
\usepackage{bm}

\newcommand{\beq}{\begin{equation}}
\newcommand{\eeq}{\end{equation}}
\newcommand{\bea}{\begin{eqnarray}}
\newcommand{\eea}{\end{eqnarray}}

\newcommand\eqn[1]{(\ref{#1})}      
\newcommand\Eqn[1]{Eq.~(\ref{#1})}  

\newcommand{\tr}{\hbox{tr}}

\newcommand{\cb}{{\bar c}}

\title{A class of nonperturbative nonlinear covariant gauges in Yang-Mills theories}

\ShortTitle{Nonlinear covariant gauges}

\author{Julien Serreau\\
        Astro-Particule et Cosmologie (APC), CNRS UMR 7164, Universit\'e Paris 7 - Denis Diderot\\ 10, rue Alice Domon et L\'eonie Duquet, 75205 Paris Cedex 13, France.\\
        E-mail: \email{serreau@apc.univ-paris7.fr}}

\abstract{We report on the recent proposal \cite{Serreau:2013ila} of a class of nonlinear covariant gauges that can be formulated as an extremization procedure which admits a simple discretization well-suited to numerical minimization techniques. This class of gauges is continuously connected to the Landau gauge and, in the ultraviolet, where one can ignore Gribov ambiguities, it reduces to the Curci-Ferrari-Delbourgo-Jarvis gauges.}

\FullConference{QCD-TNT-III-From quarks and gluons to hadronic matter: A bridge too far?,\\
		2-6 September, 2013\\
		European Centre for Theoretical Studies in Nuclear Physics and Related Areas (ECT*), Villazzano, Trento (Italy)}

\begin{document}

\section{Introduction}

The calculation of gluon and ghost correlation functions in (gauge fixed) Yang-Mills theories, both in the vacuum and at finite temperature/density, is the subject of an intense research activity. Such correlators are the basic ingredients for the calculation of (gauge invariant) observables, such as the glueball spectrum, thermodynamic quantities, or real-time transport coefficients. In this context, the Landau and Coulomb gauges are the most widely used, mainly because they can be relatively easily implemented in genuine nonperturbative approaches on the lattice \cite{Boucaud:2011ug,Maas:2011se}. For instance, the Landau gauge condition, $\partial_\mu A_\mu^g=0$, where $A_\mu^g$ is the gluon field configuration along the gauge orbit $g$, extremizes the functional (a sum over $\mu$ is understood)
\begin{equation}
\label{eq:Landau}
  {\cal F}[A,g]=\int_x\,\tr\left(A^g_\mu\right)^2,
\end{equation}
which admits a simple lattice discretization, well-suited to numerical minimization techniques.

It is of interest to investigate other possible gauges. Covariant gauges are of particular interest, in particular in the context of continuum approaches, due to their greater degree of symmetry. The simplest example beyond the Landau gauge is the class of linear covariant gauges, $\partial_\mu A_\mu^g=\Lambda$, where $\Lambda$ is a Gaussian distributed field in the Lie algebra of the gauge group. Attempts to formulate a lattice version of the latter, based on an extremization functional of the form ${\cal F}[A,g]+h[g]$,\footnote{Alternative ways to implement covariant gauges in a nonperturbative setup, not based on an extremization problem, have also been considered; see, e.g., \cite{Parrinello:1990pm,Zwanziger:1990tn,Fachin:1991pu,Henty:1996kv,Cucchieri:2008zx,Kalloniatis:2005if,vonSmekal:2008en}; see also the review \cite{Giusti:2001xf} and references therein.} have been made in Refs. \cite{Cucchieri:2009kk,Cucchieri:2010ku,Cucchieri:2011pp}. But this is doomed to failure due to a no-go theorem by Giusti \cite{Giusti:1996kf} and these proposals are in fact limited to infinitesimal gauge transformations.\footnote{This limitation concerns the possibility to describe linear covariant gauges. However, without this restriction, the proposal of \cite{Cucchieri:2009kk,Cucchieri:2010ku,Cucchieri:2011pp} can be reinterpreted as a nonlinear covariant gauge; see below.}  An alternative extremization functional, of the form $\int_x\tr(\partial_\mu A_\mu^g-\Lambda)^2$ has been considered in \cite{Giusti:1996kf,Giusti:1999im,Giusti:2000yc}. Although this presents spurious solutions, the authors argue that this can be kept under control. Another issue is that the corresponding Faddeev-Popov operator is in fact not that of linear covariant gauges. To our knowledge, this line of investigation has not been pursued further.

Another possibility is, instead, to leave out the requirement of a linear gauge but to keep the simple form of the extremization functional ${\cal F}[A,g]+h[g]$. In \cite{Serreau:2013ila}, we have proposed a class of non-linear covariant gauges that can be formulated as an extremization problem of this form which has all the good properties for a possible numerical implementation. This is a simple generalization of the proposal of \cite{Cucchieri:2009kk}, however not restricted to infinitesimal gauge transformations. Remarkably, when ignoring Gribov ambiguities at high energy, using the Standard Faddeev-Popov approach, the proposed class of gauges reduces to the known Curci-Ferrari-Delbourgo-Jarvis (CFDJ) gauges \cite{Curci:1976bt,Delbourgo:1981cm}. In this contribution, we briefly review the main lines of our proposal \cite{Serreau:2013ila}. We also comment on a continuum formulation of this class of gauges which, unlike the Faddeev-Popov procedure, consistently takes into account the Gribov ambiguities and can be formulated in terms of a local action which is perturbatively renormalizable in four dimensions.

\section{A class of non-linear covariant gauges as an extremization procedure}

In the context of an SU($N$) Yang-Mills theory, we consider the functional
\begin{equation}
  \label{eq_func}
  {\cal H}[A,\eta,g]=\int_x\,\tr\left[\left(A^g_\mu\right)^2+ \frac{g^\dagger \eta+\eta^\dagger g}2\right],
\end{equation}
where $\eta$ is an arbitrary $N\times N$ matrix field,  $A_\mu^g=gA_\mu g^\dagger+\frac i {g_0}g\partial_\mu g^\dagger$ is the gauge transform of the gluon field $A_\mu$, with $g\in$ SU($N$), and $g_0$ is the bare coupling constant. We define our gauge condition as (one of) the extrema of ${\cal H}$ with respect to $g$, which corresponds to ($a=1,\ldots,N$)
\begin{equation}
  \label{eq_eq_mot}
  \left(\partial_\mu A_\mu^{g}\right)^a=\frac{ig_0}{2}\tr\left[t^a\left(g\eta^\dagger-\eta g^\dagger\right)\right],
\end{equation}
where $t^a$ denotes the SU($N$) generators, normalized as $\tr[t^at^b]=\delta^{ab}/2$.

This can be used as a gauge condition for any $\eta$. Alternatively, we 
can average over $\eta$ with a given distribution ${\cal P}[\eta]$. Here, we choose a simple Gaussian 
distribution 
\begin{equation}
  \label{eq_p_eta}
  \mathcal P[\eta]={\cal N}\exp\left(-\frac {g_0^2}{4\xi_0}\int_x\tr\, \eta^\dagger\eta\right),
\end{equation}
with ${\cal N}$ a normalization factor. The above procedure is a simple generalization of the Landau gauge, which corresponds to the case $\eta=0$, or, equivalently, to the distribution \eqn{eq_p_eta} with $\xi_0=0$. 

The gauge fixing described here generalizes the proposal of Ref. \cite{Cucchieri:2009kk}. There, the authors considered the extrema of a similar functional as \eqn{eq_func}, where $\Lambda=-ig_0\eta$ was constrained to be an Hermitian matrix field in the Lie algebra of the gauge group, with the aim of enforcing a linear gauge condition. We see from \Eqn{eq_eq_mot} that this is in fact limited to infinitesimal gauge transformations $g\approx 1\hspace{-.13cm}1+ig_0\lambda$, for which the condition \eqn{eq_eq_mot} indeed reduces to $\partial_\mu A_\mu^{g}=\Lambda$. Here, we consider the condition \eqn{eq_eq_mot} for arbitrary $g$. Another important difference lies in the sampling \eqn{eq_p_eta} over the matrix field $\eta$ which is not restricted to the Lie algebra of the gauge group. This makes an important difference, e.g., for a continuum formulation of the corresponding gauge fixing procedure, as discussed below. For instance, the sampling measure \eqn{eq_p_eta} allows one to apply the standard Faddeev-Popov construction and leads to a relatively simple form of the effective gauge-fixed action.

To gain more insight on the gauge-fixing procedure described above, let us indeed consider the ultraviolet regime where Gribov ambiguities issues are expected to be irrelevant and the standard Faddeev-Popov procedure is justified. The Faddeev-Popov operator corresponding to the gauge condition \eqn{eq_eq_mot} reads
\begin{equation}
  \label{eq_FP_det}
 \bigg\{\partial_\mu D^{ac}_\mu[A^g]+\frac
 {g_0^2}{2}\tr\left(t^a t^cg\eta^\dagger+\eta g^\dagger t^c t^a\right)\bigg\} \delta^{(d)}(x-y)   ,
\end{equation}
where the derivatives act on the variable $x$ and where $D_\mu[A^g]$ is the standard covariant derivative in the adjoint representation evaluated at $A^g(x)$.
Introducing a Nakanishi-Lautrup field $ih$ to account for the gauge condition \eqn{eq_eq_mot} as well as ghost and antighost fields $c$ and $\cb$ to cope for the corresponding Jacobian, the Faddeev-Popov gauge-fixed action reads, for a given external field~$\eta$,
\beq
\label{eq:eff}
 S_{\rm gf}^\eta[A,c,\cb,h,g]=S_{\rm YM}[A]+S_{\rm FP}^\eta[A,c,\cb,h,g]
\eeq
with $S_{\rm YM}[A]$ the Yang-Mills action and 
\begin{equation}
  \label{eq_fp}
  \begin{split}
  S_{{\rm FP}}^\eta[A,c,\cb,h,g]=\int_x\left\{\partial_\mu\bar c^aD_\mu[A^g] c^a+ih^a\left(\partial_\mu A_\mu^g\right)^a+\frac{g_0}{2}\tr\left[\eta^\dagger R+R^\dagger\eta\right]\right\}. 
  \end{split}
\end{equation}
where\footnote{Here, $ih=ih^a t^a$ is to be seen as an Hermitian matrix field and similarly for $c$ and $\bar c$.}
\beq
 R=(h-g_0\bar c c) g.
\eeq

It is important to notice here that the effective action \eqn{eq_fp} depends separately on $A$ and $g$, not only on the combination $A^g$, which makes the standard Faddeev-Popov trick of factorizing out a volume of the gauge group inaplicable. Here, the sampling \eqn{eq_p_eta} over $\eta$ is of great help since
\beq
\label{eq:trace}
 \int {\cal D}\eta{\cal P}[\eta]\exp\left\{-\frac{g_0}{2}\int_x\tr\left[\eta^\dagger R+R^\dagger\eta\right]\right\}\propto\exp\left\{\xi_0\int_x\tr\left[R^\dagger R\right]\right\}
\eeq
does not depend explicitly on $g$ anymore.\footnote{We note that this is not true for the sampling proposed in \cite{Cucchieri:2009kk}.} The resulting gauge-fixed action is of the form $S_{\rm YM}[A]+S_{\rm FP}[A^g,c,\cb,h]$ and one can factor out the volume of the gauge group in the standard manner. Remarkably the calculation of $\tr\left[R^\dagger R\right]$ in \eqn{eq:trace} yields, after some simple algebra,
\beq
 S_{\rm gf}[A,c,\cb,h]=S_{\rm YM}[A]+S_{\rm CFDJ}[A,c,\cb,h],
\eeq
where 
\begin{equation}
  \label{eq_action_CFDJ}
  \begin{split}
  S_{\rm CFDJ}[A,c,\cb,h]=\int_x\bigg\{ \partial_\mu \cb^aD_\mu c^a+ih^a \partial_\mu A_\mu^a
  +\xi_0 \bigg[\frac {(h^a)^2}2\!-\!\frac {g_0}2 f^{abc}ih^a\cb^b c^c\!-\!\frac {g_0^2}4 \left( f^{abc}\cb^b c^c\right)^2\bigg]\bigg\}    
  \end{split}
\end{equation}
is nothing but the Curci-Ferrari-Delbourgo-Jarvis gauge-fixing action \cite{Curci:1976bt,Delbourgo:1981cm}. Thus the gauge fixing discussed here provides a nonperturbative generalization of the CFDJ gauge that may be implementable on the lattice; see below. 

The CFDJ gauges are known to possess various good properties. For instance, they are perturbatively 
renormalizable in four dimensions. Also, they have a nilpotent BRST symmetry and are thus unitary. 
However, they have Gribov ambiguities, just 
as the Landau gauge. In principle this would not be a problem for lattice calculations 
as one may easily select a particular copy, as done in the so-called 
minimal Landau gauge. 

\section{Lattice formulation}

Let us briefly discuss the possible lattice formulation of the gauge-fixing described here, following the lines of \cite{Cucchieri:2009kk}. Introducing the SU($N$) lattice link variable $U_\mu(x)=\exp\left\{-iag_0A_\mu(x)\right\}$ and the rescaled matrix field $M(x)=a^2g_0^2\eta(x)/2$, the simplest discretization of the functional \eqn{eq_func} reads
\beq
\label{eq:minfunc}
 {\cal H}_{\rm latt.}[U,M,g]={\rm Re}\,\,{\rm tr}\left\{-\sum_{x,\mu} U^g_\mu(x)+\sum_xM^\dagger(x)g(x)\right\},
\eeq
with the gauge transformed link variable $U^g_\mu(x)=g(x)U_\mu(x)g^\dagger(x+\hat\mu)$, where $\hat\mu$ denotes a lattice link in the direction $\mu$. Defining
\beq
 \mathbb{A}^a_\mu(x)=2\tr\left[t^a\frac{U_\mu^\dagger(x)-U_\mu(x)}{2i}\right]\qquad{\rm and}\qquad \nabla\cdot\mathbb{A}^a(x)=\sum_\mu\left[\mathbb{A}^a_\mu(x+\hat\mu)-\mathbb{A}^a_\mu(x)\right],
\eeq
the extrema of the functional \eqn{eq:minfunc} satisfy the lattice gauge condition
\beq
 \left(\nabla\cdot\mathbb{A}^g\right)^a=i\,\tr\left[t^a\left(gM^\dagger-Mg^\dagger\right)\right].
\eeq

The first term on the right hand side of \Eqn{eq:minfunc} is the usual discretized version of the Landau gauge extremization functional \eqn{eq:Landau}. Its essential property is that it is linear in the gauge transformation matrix $g(x)$ at each lattice site $x$, which permits the use of powerful numerical minimization techniques \cite{Boucaud:2011ug}. This property is obviously true for the whole functional \eqn{eq:minfunc}, which suggests that similar minimization techniques can be employed in that case as well. 

Once a given extremum (minimum) has been obtained for each configuration of the link variables $U_\mu(x)$ and of the noise field $M(x)$, one performs the average over the former with the (discretized) Yang-Mills action and over the latter with a given weight
\beq
 {\cal P}_{\rm latt.}[M]=\prod_xp_{\rm STT}\big(M(x)\big).
\eeq
Our proposal \cite{Serreau:2013ila} corresponds to
\beq
\label{eq:STT}
 p_{\rm STT}(M)=\exp\left\{-\frac{1}{\xi_0g_0^2}{\rm tr}\left[M^\dagger M\right]\right\}=\prod_{a=0}^N\exp\left\{-\frac{|M_a|^2}{2\xi_0g_0^2}\right\}.
\eeq
where, in the second equality, we introduced the decomposition 
\beq
 M=\frac{M_0}{\sqrt{2N}}1\hspace{-.13cm}1 +\sum_{a=1}^NM_at^a.
\eeq
For comparison, the proposal of Cucchieri, Mendes and Santos \cite{Cucchieri:2009kk} corresponds to the following averaging over the noise field $M(x)$, in the present notations,
\beq
 p_{\rm CMS}(M)=\delta^{(2)}(M_0)\times\prod_{a=1}^N\delta({\rm Re}M_a)\exp\left\{-\frac{({\rm Im} M_a)^2}{2\xi_0g_0^2}\right\}.
\eeq
Although the ability of the proposal of \cite{Cucchieri:2009kk} to describe linear covariant gauges is limited to infinitesimal gauge transformations, we believe that its actual numerical implementation is not. In practice, the minimization algorithm implemented in \cite{Cucchieri:2009kk,Cucchieri:2010ku,Cucchieri:2011pp} shows good convergence properties. We see no reason to expect the different sampling on $M$ to be an issue and it would thus be of great interest to investigate whether similar numerical methods apply to the present proposal.

\section{Continuum formulation: lifting Gribov ambiguities}

The class of gauges discussed here has Gribov ambiguities that must be taken into account away from the high momentum perturbative regime. This would, in principle, be easy to do on the lattice, e.g., by picking just one copy (one minimum of the extremization functional) per gauge orbit, as usually done in the so-called minimal Landau gauge. This procedure, however, has no direct continuum formulation and cannot be implemented as such in continuum approaches. In \cite{Serreau:2012cg} (see also \cite{Serreau:2013ds} for a brief description) we have proposed, in the context of the Landau gauge, an alternative approach, based on averaging over the various Gribov copies of each gauge orbit, which can be formulated in terms of a local renormalizable action in four dimensions. Remarkably, this effectively results in a simple massive extension of the standard Faddeev-Popov gauge-fixed action which is a particular case of the Curci-Ferrari model \cite{Curci:1976bt}.\footnote{More precisely, the averaging procedure over Gribov copies produces an effective bare mass for gluons which has to be sent to zero at the end of any calculation, together with the (continuum) limit of vanishing bare coupling. This can be done by keeping the renormalized mass finite, in which case the gauge-fixed action turns out to be perturbatively equivalent to the Curci-Ferrari model for what concerns the calculation of gluon and ghost correlators. We mention that the relation between the Curci-Ferrari mass term and the average over Gribov copies has been noticed earlier in a slightly different context in \cite{vonSmekal:2008en}.} This provides, for the first time, a first principle link between this model and Yang-Mills theories. This is of great interest in relation with the recent observation \cite{Tissier:2010ts,Tissier:2011ey} that a simple one-loop perturbative calculation in the Curci-Ferrari model reproduces the lattice results for the vacuum two-point gluon and ghost correlators, with remarkable accuracy, down to deep infrared momenta. This has been recently extended to higher correlation functions in the vacuum \cite{Pelaez:2013cpa} and to the study of two-point correlators at finite temperature \cite{Reinosa:2013twa}.

In \cite{Serreau:2013ila}, we have generalized this approach of averaging over Gribov copies to the class of nonlinear covariant gauges discussed here. We have shown that, again, this can be formulated as a local action which is perturbatively renormalizable in four dimension and we have computed the five independent renormalization factors at one-loop order. In that case, the effective theory is not a simple massive extension of the CFDJ action. It also includes a set of replicated scalar, ghost and antighost as well as Nakanishi-Lautrup fields which do not decouple (as in the case of the Landau gauge). It is interesting to perform a one-loop calculation of the gluon and ghost two-point correlators in this class of gauges, for instance to study their dependence on the gauge fixing parameter $\xi_0$. This is work under progress. We hope this will stimulate the lattice community to try to implement our proposal \cite{Serreau:2013ila} in actual numerical calculations. This would allow one to study Yang-Mills correlators in a class of gauges continuously connected to the Landau gauge.

\section*{Acknowledgements}

I would like to thank the organizers of this stimulating workshop. I also thank my collaborators U. Reinosa, M. Tissier, A. Tresmontant and N. Wschebor as well as J. Greensite, M. Huber, A. Maas and L. von Smekal  for interesting discussions.


\begin{thebibliography}{99}

\bibitem{Serreau:2013ila}
  J.~Serreau, M.~Tissier and A.~Tresmontant,
  arXiv:1307.6019 [hep-th].
  
\bibitem{Boucaud:2011ug}
  P.~.Boucaud, J.~P.~Leroy, A.~L.~Yaouanc, J.~Micheli, O.~Pene and J.~Rodriguez-Quintero,
  Few Body Syst.\  {\bf 53} (2012) 387.

\bibitem{Maas:2011se}
  A.~Maas,
  Phys.\ Rept.\  {\bf 524} (2013) 203.

\bibitem{Parrinello:1990pm}
  C.~Parrinello and G.~Jona-Lasinio,
  Phys.\ Lett.\ B {\bf 251} (1990) 175.

\bibitem{Zwanziger:1990tn}
  D.~Zwanziger,
  Nucl.\ Phys.\ B {\bf 345} (1990) 461.

\bibitem{Fachin:1991pu}
  S.~Fachin and C.~Parrinello,
  Phys.\ Rev.\ D {\bf 44} (1991) 2558.

\bibitem{Henty:1996kv}
  D.~S.~Henty {\it et al.}  [UKQCD Collaboration],
  Phys.\ Rev.\ D {\bf 54} (1996) 6923.

\bibitem{Cucchieri:2008zx}
  A.~Cucchieri, A.~Maas and T.~Mendes,
  Comput.\ Phys.\ Commun.\  {\bf 180} (2009) 215.

\bibitem{Kalloniatis:2005if}
  A.~C.~Kalloniatis, L.~von Smekal and A.~G.~Williams,
  Phys.\ Lett.\ B {\bf 609} (2005) 424.

\bibitem{vonSmekal:2008en}
  L.~von Smekal, M.~Ghiotti and A.~G.~Williams,
  Phys.\ Rev.\ D {\bf 78} (2008) 085016.
  
\bibitem{Giusti:2001xf}
  L.~Giusti, M.~L.~Paciello, C.~Parrinello, S.~Petrarca and B.~Taglienti,
  Int.\ J.\ Mod.\ Phys.\ A {\bf 16} (2001) 3487.

\bibitem{Cucchieri:2009kk}
  A.~Cucchieri, T.~Mendes and E.~M.~S.~Santos,
  Phys.\ Rev.\ Lett.\  {\bf 103} (2009) 141602.

\bibitem{Cucchieri:2010ku}
  A.~Cucchieri, T.~Mendes and E.~M.~d.~S.~Santos,
  PoS QCD {\bf -TNT09} (2009) 009.
  
\bibitem{Cucchieri:2011pp}
  A.~Cucchieri, T.~Mendes, G.~M.~Nakamura and E.~M.~S.~Santos,
  PoS FACESQCD {\bf } (2010) 026.

\bibitem{Giusti:1996kf}
  L.~Giusti,
  Nucl.\ Phys.\ B {\bf 498} (1997) 331.

\bibitem{Giusti:1999im}
  L.~Giusti, M.~L.~Paciello, S.~Petrarca and B.~Taglienti,
  Phys.\ Rev.\ D {\bf 63} (2001) 014501.
  
\bibitem{Giusti:2000yc}
  L.~Giusti, M.~L.~Paciello, S.~Petrarca, C.~Rebbi and B.~Taglienti,
  Nucl.\ Phys.\ Proc.\ Suppl.\  {\bf 94} (2001) 805.

\bibitem{Curci:1976bt}
  G.~Curci, R.~Ferrari,
  Nuovo Cim.\ A {\bf 32} (1976) 151;
  Nuovo Cim.\ A {\bf 35} (1976) 1.

\bibitem{Delbourgo:1981cm}
  R.~Delbourgo and P.~D.~Jarvis,
  J.\ Phys.\ A {\bf 15} (1982) 611.

\bibitem{Serreau:2012cg}
  J.~Serreau and M.~Tissier,
  Phys.\ Lett.\ B {\bf 712} (2012) 97.

\bibitem{Serreau:2013ds}
  J.~Serreau,
  PoS ConfinementX {\bf } (2012) 072.

\bibitem{Tissier:2010ts}
  M.~Tissier and N.~Wschebor,
  Phys.\ Rev.\ D {\bf 82} (2010) 101701.

\bibitem{Tissier:2011ey}
  M.~Tissier and N.~Wschebor,
  Phys.\ Rev.\ D {\bf 84} (2011) 045018.

\bibitem{Pelaez:2013cpa}
  M.~Pel\'aez, M.~Tissier and N.~Wschebor,
  Phys.\ Rev.\ D {\bf 88} (2013) 125003.

\bibitem{Reinosa:2013twa}
  U.~Reinosa, J.~Serreau, M.~Tissier and N.~Wschebor,
  arXiv:1311.6116 [hep-th].
  
\end{thebibliography}
\end{document}